\documentclass[showpacs,twocolumn,preprintnumbers,amsmath,amssymb,epsf,superscriptaddress]{revtex4-1}
\usepackage{txfonts}
\usepackage{graphicx}
\usepackage{lipsum}
\usepackage{nicefrac}
\usepackage{hyperref}
\usepackage{color}
\usepackage[squaren]{SIunits}
\newcommand{\be}{\begin{equation}}
\newcommand{\ee}{\end{equation}}
\newcommand{\ber}{\begin{eqnarray}}
\newcommand{\eer}{\end{eqnarray}}
\newcommand{\de}{\end{equation*}}
\newcommand{\cer}{\begin{eqnarray*}}
\newcommand{\der}{\end{eqnarray*}}
\begin{document}
\title{A polarization gadget with two quarter wave plates: Application to Mueller Polarimetry}
\author{Salla Gangi Reddy$^*$}
\affiliation{Physical Research Laboratory, Navarangpura, Ahmedabad, India - 380009.}
\author{Shashi Prabhakar}
\affiliation{Physical Research Laboratory, Navarangpura, Ahmedabad, India - 380009.}
\author{Chithrabhanu P}
\affiliation{Physical Research Laboratory, Navarangpura, Ahmedabad, India - 380009.} 
\author{R. P. Singh}
\affiliation{Physical Research Laboratory, Navarangpura, Ahmedabad, India - 380009.}

\author{R. Simon}
\affiliation{The Institute of Mathematical Sciences, Chennai, India - 600113. \\ $^*$sgreddy@prl.res.in}

\date{\today}
\vspace{-10mm}
\begin{abstract}
We show that there are number of ways to transform an arbitrary polarization state to another with just two quarter wave plates (QWP). We have verified this geometrically using the trajectories of the initial and final polarization states corresponding to all the fast axis orientations of a QWP on the Poincar\'e sphere. The exact analytical expression for the locus of polarization states has also been given that describes the trajectory. An analytical treatment of the equations obtained through matrix operations corresponding to the transformation supports the geometrical representation. This knowledge can be used to obtain the Mueller matrix by just using quarter wave plates which has been shown experimentally by exploiting projections of the output states on the input states. 
\end{abstract}
\maketitle

The controlled transformation of one polarization state to another is of great importance in polarization optics \cite{gangi1,gangi2,gangi3,mueller1}. Quarter wave plates (QWP) and half wave plates (HWP) are the basic tools to transform the polarization states. QWP changes the linearly polarized light to elliptically or circularly polarized light and vice versa where as HWP changes the azimuth angle of the linearly polarized light \cite{gangi4}. All the polarization states can be represented on the Poincar\'e sphere geometrically and points with unit degree of polarization are on the surface that form a SU(2) group \cite{gangi5}. The state with an arbitrary polarization can be converted to any other with a group of wave plates. However, one needs to have minimum three components to realize all the SU(2) operations \cite{gangi6,gangi7} as it is a three generator group. We proposed a novel method for Mueller polarimetry with two universal SU(2) polarization gadgets consisting of two QWPs and one HWP, one for the generation of input states and another for projecting the output state \cite{gangi8}. Recently, De Zela has proposed that two QWPs can be used for the arbitrary transformation of polarization states. It was emphasized that although the transformation process was unique, however, two QWPs couldn't transform the polarization states to corresponding orthogonal states \cite{gangi9}. Here, we show that the transformation of all polarization states from one to another is possible with two QWPs \textit{including orthogonal states} and this is not unique. We show a number of ways to transform the states using two QWPs. However, \textit{it is not possible to realize all SU(2) operations with this gadget such as rotating all the states by same amount} which is possible with a universal SU(2) gadget.  
 
 First, we consider a polarization state with polar and azimuthal angles as ($\phi, \alpha$) on the Poincar\'e sphere with Stokes parameters $(1,q,u,v)^T$ 
 
\begin{center}
\begin{equation}
\label{Mueller1}
S = \begin{bmatrix}
       1            \\[0.3em]
       \cos(\phi)\cos(\alpha)        \\[0.3em]
       \cos(\phi)\sin(\alpha)         \\[0.3em]
       \sin(\phi)          
            \end{bmatrix}. 
\end{equation}
\end{center}  
The action of any optical component on a given polarization state can be studied simply by operating the corresponding Mueller matrix on the Stokes vector of the polarization state. Here, QWP is our point of interest. The Mueller matrix of a QWP with the fast axis orientation $\theta$ is \cite{gangi5}

\begin{figure}[htb]
 \begin{center}
 \includegraphics[height=1.0in]{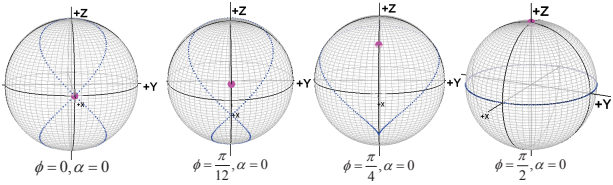} 
  \caption{(Colour online) Quarter wave plate action on four different polarization states (shown by maroon colour spots on the spheres) and showing the increase in asymmetry of the dumbbells in the trajectories at $\alpha=0$.}
  \label{fig:mueller1} 
 \end{center}
 \end{figure}

\begin{center}
\begin{equation}
\label{Mueller2}
 Q_{\theta} = \begin{bmatrix}
        1 \hspace{0.001cm}&0 \hspace{0.001cm}& 0 \hspace{0.001cm}& 0 \\[0.2em]
        0 \hspace{0.001cm}&\cos^2(2\theta) \hspace{0.001cm}& \sin(2\theta)\cos(2\theta) \hspace{0.001cm}&-\sin(2\theta) \\[0.2em]
        0 \hspace{0.001cm}&\sin(2\theta)\cos(2\theta) \hspace{0.001cm}&\sin^2(2\theta) \hspace{0.001cm}&-\cos(2\theta) \\[0.2em]
        0 \hspace{0.001cm}&\sin(2\theta) \hspace{0.001cm}&-\cos(2\theta) \hspace{0.001cm}&0
        \end{bmatrix}.
\end{equation}
\end{center}
When the QWP acting on the state given in Eq. \ref{Mueller1}, the output state becomes  $(1,q',u',v')^T$
\begin{equation}
S' = \begin{bmatrix}
       1            \\[0.3em]
      \cos(2\theta)\cos(\phi)\cos(2\theta-\alpha) - \sin(2\theta)\sin(\phi)     \\[0.3em]
       \sin(2\theta)\cos(\phi)\cos(2\theta-\alpha)+\cos(2\theta)\sin(\phi)         \\[0.3em]
       \cos(\phi)\sin(2\theta-\alpha)        
            \end{bmatrix}. 
            \label{Mueller3}
\end{equation}   
By removing the dependence of $\theta$ in Eq. \ref{Mueller3}, one can get the locus of all polarization states corresponding to all the fast axis orientations of QWP. This gives the trajectory of polarization states on the Poincar\'e sphere which is given by 

\begin{equation}
\label{Mueller4}
q'-a^2\cos(\phi)\cos(\alpha)-\sqrt{1-a^2} \{a \cos(\phi)\sin(\alpha)-\sin(\phi)\} = 0 
\end{equation} where
\begin{equation}
\label{Mueller5}
 a = \frac{u'-\cos(\phi)\sin(\alpha)}{\sin(\phi)+v'}.
\end{equation}
At $\alpha=0$, the trajectory becomes 
 
 \begin{equation}
 \label{Mueller6}
q'+\frac{u'.v'}{\sqrt{\cos^2(\phi)-v'^2}}-\cos(\phi) = 0.
\end{equation}
Figure \ref{fig:mueller1} shows the geometrical representation of the trajectory of the polarization states corresponding to different fast axis orientations of the QWP for various values of $\phi$ at $\alpha=0$. As shown in the figure, these trajectories are dumbbell in shape on the surface of the Poincar\'e sphere. The lobes of a dumbbell will appear as circles when we observe them in a 2-D plane. The end points of two lobes of dumbbell of a given trajectory are always antipodal (diagonally opposite points) on the sphere. The two lobes are perfectly symmetric (both are of same area on the surface of the Poincar\'e sphere) at $\phi=0$ and become asymmetric for non-zero $\phi$ values. The asymmetry increases with the increase in $\phi$ from $0^\circ$ to $90^\circ$ and finally one lobe disappears when the orientation of fast axis of the QWP exceeds 45$^\circ$. The trajectory becomes a circle on the Poincar\'e sphere at $\phi=90^\circ$. If $\phi$ is $90^\circ$, the state $S$ becomes a completely circularly polarized light which can be converted to plane polarized light by any fast axis orientation of the QWP. This gives a circular trajectory for all the output states in the equatorial plane (plane polarized) of the Poincar\'e sphere. The similar behaviour can be seen in the opposite plane if we change $\phi$ from $90^\circ$ to $180^\circ$. 

 For studying the transformation of one polarization state to another, we have considered two arbitrary polarization states $S_1$ and $S_2$ on the Poincar\'e sphere with Stokes parameters $(1, q_i, u_i, v_i)^T$ (where $i = 1,2$). For completely polarized light, we have
\be
q_i^2+u_i^2+v_i^2=1.
\ee
The transformation of one state to another with two QWPs having Mueller matrices $Q_{\theta1}$ and $Q_{\theta2}$, where $\theta1$ and $\theta2$ are the fast axis orientations of two QWPs, can be written as 
\be
\label{eq}
S_2 = Q_{\theta2} \hspace{0.1cm} Q_{\theta1} \hspace{0.1cm} S_1. 
\ee   
One needs to find the two fast axis orientations to transform one state to another. To the best of our knowledge, there is no direct formula for the determination of fast axis orientations of two QWPs in terms of Eulers angles of input and output states as in the case of universal SU(2) gadget \cite{gangi7}. Here, we use the geometrical representation for finding these orientations and the details are as follows.          
 
It is well known that one can realize the inverse operation of QWP with itself as
\be
\label{eq2}
Q^{-1}_\theta = Q_{\theta+\frac{\pi}{2}}.
\ee
Equation \ref{eq} can also be written as 
\be
Q_{\theta2}^{-1}.S_2 = Q_{\theta1}.S_1.
\ee
 Using the Eq. \ref{eq2}, one can write the above equation as 
 \be
 \label{reddy}
Q_{\theta2}.S_2 = Q_{\theta1}.S_1; \hspace{0.3cm} S'_2 = S'_1.
 \ee  
From the above equation, it is clear that if two output Stokes vectors obtained after the action of QWP are equal, the two states can be transformed to one another using two QWPs. 

\begin{center}
\begin{figure}[htb]
  \includegraphics[height=1.0in]{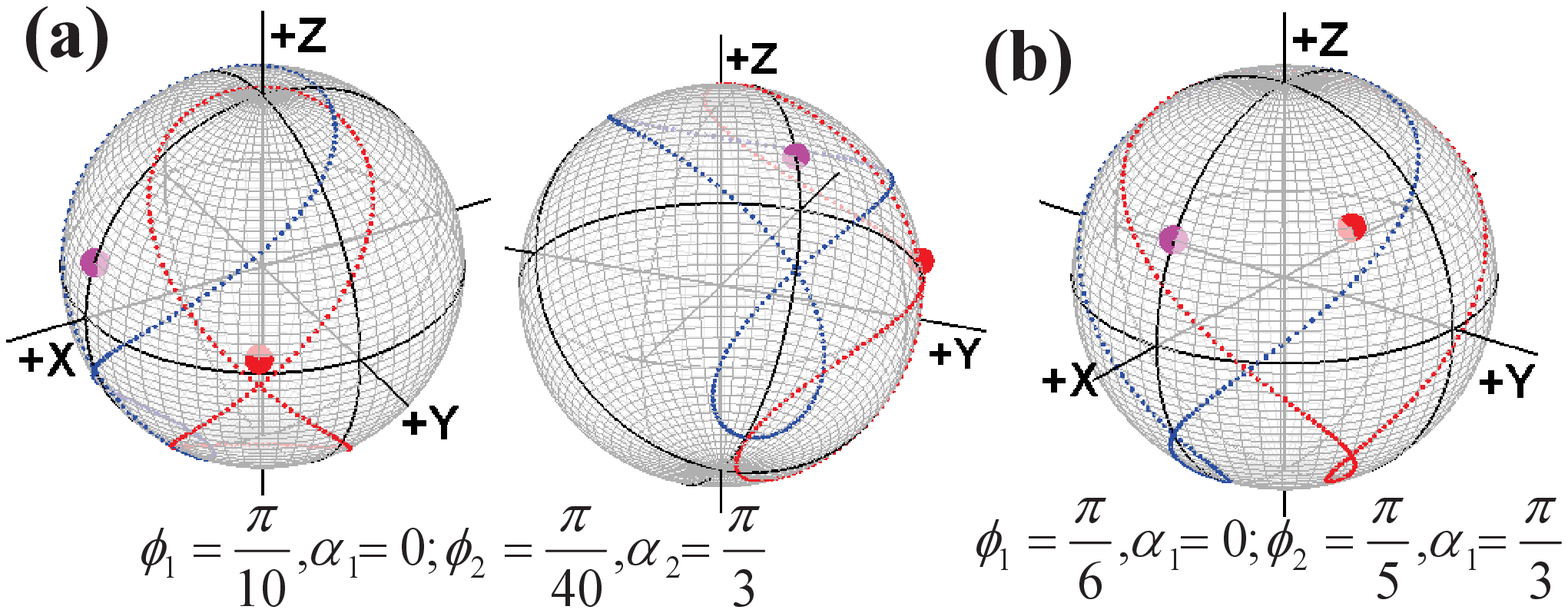} 
  \caption{(Colour online)  The points of intersection of two trajectories followed by the two non-orthogonal initial polarization states (a) trajectories intersected at four points. (b) trajectories intersected at two points.}
  \label{fig:mueller2} 
 \end{figure}
  \end{center}

\begin{figure}[htb]
 \begin{center}
 \includegraphics[height=2.0in]{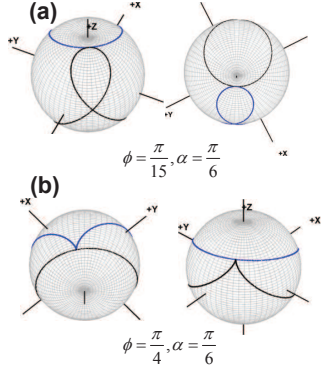} 
  \caption{(Colour online) The points of intersection of two trajectories followed by the two orthogonal initial polarization states and both the trajectories intersected at two points.}
  \label{fig:mueller3} 
 \end{center}
 \end{figure}

To determine the required fast axis orientations, we have plotted the trajectories given by Eq. \ref{Mueller4} for the two initial states on the Poincar\'e sphere. The points of intersection (where the two output Stokes vectors are same) of these two trajectories give solutions to transform one state to another. The number of points of intersection is equal to the number of possibilities for the transformation. We have plotted the states $S'_1$ and $S'_2$ on the Poincar\'e sphere as functions of $\theta1$ and $\theta2$ respectively as shown in Fig \ref{fig:mueller2} (a) and (b). We observe that the number of solutions depends on the selection of input and output polarization states. The minimum number of solutions we have obtained to transform an arbitrary state to another are two. One can see from the figure that there are four solutions for one set of states (Fig \ref{fig:mueller2} (a)) while two for another set (Fig \ref{fig:mueller2} (b)). We have shown two different planes of Poincar\'e sphere in Fig \ref{fig:mueller2} (a) to visualize all the solutions for a single set of states correspond to $S_1$ ($\phi_1 = \pi/10$, $\alpha_1=0$) and $S_2$ ($\phi_2 = \pi/40$, $\alpha_2 = \pi/3$). The second set of states shown in Fig \ref{fig:mueller3} correspond to $\phi_1 = \pi/6$, $\alpha_1=0$ ($S_1$); $\phi_2 = \pi/5$ and $\alpha_2 = \pi/3$ ($S_2$).

 Next, we show that the action of two QWPs can transform one state to its orthogonal state which is in contradiction to Dezela's conclusion \cite{gangi9}. The two polarization states $S_1$ and $S_2$ with Stokes parameters $(1, q_i, u_i, v_i)^T$ are said to be orthogonal to one another if and only if 
\be
 S_1 . S_2 =0 \hspace{0.2cm} \rm or \hspace{0.2cm} 1 + q_1 q_2 + u_1 u_2 + v_1 v_2 = 0. 
\ee
Now we consider two orthogonal states with Stokes vectors $(1, q_1, u_1, v_1)^T$ and $(1, -q_1, -u_1, -v_1)^T$ and follow the same procedure described above. The geometrical representation shows that one can transform two arbitrary orthogonal states into one another by two ways as the loci are touching at two points on the Poincar\'e sphere. Figure \ref{fig:mueller3} (a) shows two different planes of Poincar\'e sphere to show all solutions corresponds to $\phi = \pi/15$ and $\alpha = \pi/6$ (a) and \ref{fig:mueller3} (b) corresponds to $\phi = \pi/4$ and $\alpha = \pi/6$. 

Thus, one can determine the fast axis orientations of two QWPs for converting one state to another simply by plotting the two trajectories and locating the points of intersection. Infinite number of combinations are possible to convert right circularly polarized (RCP) light to left circularly polarized (LCP) light and vice versa. This is due to the fact that QWP at any fast axis orientation converts LC polarization to plane polarized light and correspondingly another fast axis orientation exists to convert any plane polarized light to RC polarized light. So, one can convert LCP and RCP light to one another in infinite number of ways using two QWPs.

For the analytical support to our geometrical representation, we have substituted the  Mueller matrices for two QWPs in Eq. \ref{eq} with different fast axes orientations ($\theta1$, and $\theta2$) and obtained the relationship between initial and final Stokes vectors as
\ber
\label{Mueller7}
 q' &=& a_1q+b_1u+c_1v;  \label{g1} \\   u' &=& a_2q+b_2u+c_2v; \label{g2} \\  v' &=& a_3q+b_3u+c_3v \label{g3}.
\eer
with

\ber
 a_{1} &=& \cos(2\theta1)\cos(2\theta2)\cos(2\theta)-\sin(2\theta1)\sin(2\theta2), \nonumber \\
 b_{1} &=& \cos(2\theta1)\sin(2\theta2)\cos(2\theta)+\sin(2\theta1)\cos(2\theta2), \nonumber \\ 
 c_{1} &=& \cos(2\theta1)\sin(2\theta), \nonumber \\
 a_{2} &=& \sin(2\theta1)\cos(2\theta2)\cos(2\theta)+\cos(2\theta1)\sin(2\theta2), \nonumber \\
 b_{2} &=& \sin(2\theta1)\sin(2\theta2)\cos(2\theta)-\cos(2\theta1)\cos(2\theta2), \nonumber \\  
 c_{2} &=& \sin(2\theta1)\sin(2\theta),  \nonumber \\
 a_{3} &=& \cos(2\theta2)\sin(2\theta),\nonumber \\ 
 b_{3} &=& \sin(2\theta2)\sin(2\theta), \nonumber \\
 c_{3} &=& -\cos(2\theta)
\eer
where $\theta=\theta1-\theta2$.
Equations \ref{g1}, \ref{g2}, \ref{g3} relate the output Stokes vector with the input Stokes vector and having two variables $\theta1$ and $\theta2$. The solutions of these three equations for $\theta1$ and $\theta2$, are not unique as number of variables are less than the number of equations. This is consistent with our results obtained by the geometrical representation of states.

We can use the transformation of one state to another with the two QWPs demonstrated above to find the Mueller matrix of an optical element by using the method described in reference \cite{gangi8}. However, we use a polarization gadget consisting of two QWPs instead of a universal SU(2) gadget that reduces the number of wave plates by two and consequently the errors. We provide below a brief description of the method used and the corresponding results for the free space.

When a polarization state of Stokes vector ($S$) passes through an optical component, the output state $(S')$ can be written as 
\begin{equation} \label{Ss}
  S' =  MS
\end{equation} where $S$ and $S'$ are four element column vectors and M is a $4 \times 4$ matrix defined as Mueller matrix. We form two matrices $\Omega$ and $\Omega'$ by arranging the four input and output Stoke's vectors as four columns 
\begin{equation}
  \Omega=[S_{1}~S_{2}~S_{3}~S_{4}], \hspace{1cm} \Omega'=[S'_{1}~S'_{2}~S'_{3}~S'_{4}].
\end{equation}
Using Eq. (\ref{Ss}), the relationship between $\Omega$ and $\Omega'$ can be written as
\begin{equation}
 \Omega' =  M\Omega.
\end{equation}
 
The four input states forming $\Omega$, must be on the surface of the Poincar\'e sphere as we are working with the non-depolarizing wave plates. The projection of each output state $S'_{1}, ~S'_{2}, ~S'_{3}, ~S'_{4}$ on input states $S_{1}, ~S_{2}, ~S_{3}, ~S_{4}$ gives the 16 non-negative real numbers which form the elements of projection matrix $\Lambda$. These elements are given by
\begin{equation}
\label{pro}
\Lambda_{ij} = \frac{1}{2} (S_{j})^T S'_{i} = \frac{1}{2}\sum_{\alpha=1}^{4} S_{j}^{\alpha}S_{i}^{'\alpha}
\end{equation} 
where $i,j=$ 1 to 4 and $S_i^\alpha$ denotes the $\alpha^{th}$ element of $i^{th}$ Stoke's vector and $\Lambda_{ij}$ is the projection of $S'_i$ state on $S_j$ state. Therefore, the projection matrix 
\begin{equation}
\Lambda=\frac{1}{2} \Omega^{T} \Omega' = \frac{1}{2} \Omega^{T} M \Omega .
\end{equation}
Mueller matrix $M$ can be written as \cite{gangi8}
\begin{equation} \label{mueller}
  M=2 (\Omega^T)^{-1} \Lambda \Omega^{-1}.
\end{equation}

To obtain Mueller matrix, one needs to choose four input states and determine the corresponding projection matrix. Here, we have considered the four vertices of a regular tetrahedron on the Poincar\'e sphere as input states whose Stokes vectors are 
 
\begin{equation}
\Omega=   \left[\begin{array}{r r r r}
        1.000   &   1.000  &  1.000 &  1.000 \\  
      1.000   &  -0.333  &   -0.333 & -0.334 \\  
       0.000   & 0.943  &   -0.472 & -0.471 \\  
       0.000  & 0.000  &  0.816  & -0.816
     \end{array}\right].
\end{equation}

We have generated these four input states from the vertically polarized laser light with two QWPs whose fast axis orientations ($\theta1$, $\theta2$) are determined numerically by following the procedure demonstrated above. These angles have been shown in Table (1). In order to project the output state on the four input states, we have used another set of two QWPs. The fast axis orientations used for the projections are obtained simply by adding 90$^\circ$ to the angles used while generating them with respect to the polarization of the laser beam.  

\begin{center}
 \begin{table}
   \label{tab:gan}
  \begin{minipage}[b]{1\linewidth}        \centering
 \renewcommand{\arraystretch}{1.2}
  \begin{tabular}{|c|c|c|} \hline
\hspace{0.1cm} State $S_i$ \hspace{0.1cm} & \hspace{0.1cm} $\theta1$  \hspace{0.1cm}  & \hspace{0.1cm} $\theta2 $   \hspace{0.1cm}      \\ \hline	 

 $S_{1}$ &   0$^\circ$    & 0 $^\circ$      \\ \hline
 
 $S_{2}$ & 27.48$^\circ$      & 27.28$^\circ$      \\ \hline
 
$S_{3}$ & 16.98$^\circ$       &  66.07$^\circ$     \\ \hline

$S_{4}$ & 107.25$^\circ$  & 156.42$^\circ$  \\ \hline
\end{tabular}                                
\caption{ Angles of rotation for two QWPs to generate the four vertices of a regular tetrahedrons from the vertically polarized laser beam.} 
\end{minipage}
\end{table}  
\end{center}

\begin{figure}[htb]
 \begin{center}
 \includegraphics[height=1.2in]{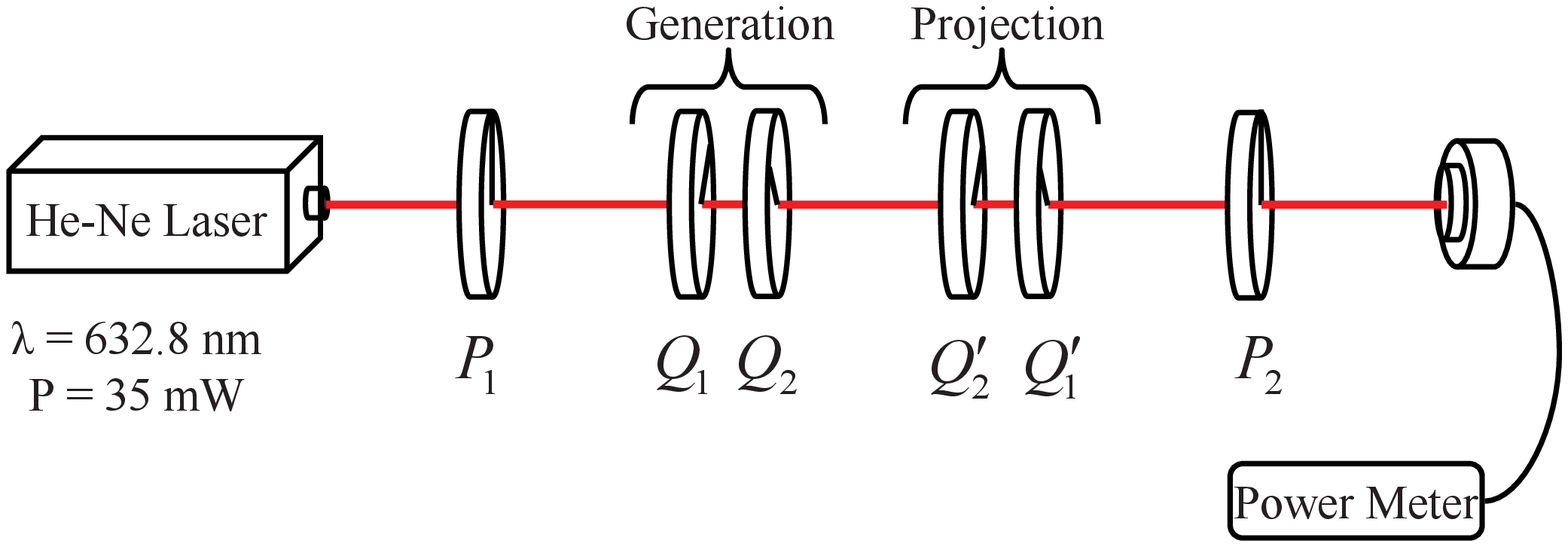} 
  \caption{(Colour online) The experimental set up to determine the Mueller matrix of an optical component. Q - QWP, P - Polarizer.}
  \label{fig:mueller8} 
 \end{center}
 \end{figure}

The experimental set up used to generate the four input states and to determine the projection matrix is shown in Fig. \ref{fig:mueller8}. A He-Ne laser having wavelength 632.8 nm and vertical polarization is used for this study. Laser beam is allowed to pass through a polarizer $P_1$ to increase the ratio of vertical to horizontal polarization of the laser and to set the reference for fast axis of all the polarizing elements. The laser beam is allowed to pass through a two component gadget ($Q_1, Q_2$) to generate the required input states using the angles shown in Table (1). In order to project the output state on the four input states, we have used another two component gadget ($Q'_1, Q'_2$) along with a polarizer $P_2$ (whose fast axis kept parallel to $P_1$) and followed the procedure described in reference \cite{gangi8}. We have used an optical multimeter (PM 100 Thorlab) of resolution 1 nW and the QWPs are from Castech. We have aligned the fast axis orientations of all QWPs by putting the polarizer $P_2$ orthogonal to $P_1$ and observing the minimum power by the detector. We have determined the projections of four output states on the input states given by the two sets of QWPs. These 16 non-negative and real elements form a projection matrix that provides us the Mueller matrix by Eq. \ref{mueller}. We have chosen free space as a sample. We have determined all projection elements theoretically by using Jones vector analysis. The theoretical and experimental projection and Mueller matrices are given below.

\ber
     \Lambda_{exp}=   \left[\begin{array}{r r r r}
        1.000   & 0.326   & 0.338  &  0.333 \\
    0.334  &  1.003   & 0.332  &  0.327 \\
    0.323  &  0.336  &  1.012   & 0.328  \\
    0.325   & 0.340 &   0.329   & 1.011   
     \end{array}\right], \nonumber \\
     \Lambda_{theo}=   \left[\begin{array}{r r r r}
        1   &   0.333  & 0.333 &  0.333 \\  
      0.333  & 1  &   0.333& 0.333 \\  
       0.333  & 0.333  &   1 & 0.333 \\  
       0.333  & 0.333  & 0.333  & 1
     \end{array}\right],  \nonumber \\
M_{exp}=   \left[\begin{array}{r r r r}
        1.000   &   -0.008  &  -0.002 &  0.001 \\  
      -0.001   &  1.010  &   -0.012 & 0.005 \\  
       -0.004   & 0.017  &   1.011 & -0.004 \\  
       -0.005  & 0.003  & -0.010  & 1.021
     \end{array}\right], \nonumber \\
     M_{theo}=   \left[\begin{array}{r r r r}
        1   &   0  & 0 &  0 \\  
      0   & 1  &   0& 0 \\  
       0  & 0  &   1 & 0 \\  
       0  & 0  & 0  & 1
     \end{array}\right]. \nonumber
\eer

From the matrices, it is clear that our experimental results are in good agreement with the theory. The maximum error obtained in the individual elements of the Mueller matrix is 0.021.

In conclusion, we have shown that there are a number of ways to transform an arbitrary polarization state to another with a polarization gadget containing two QWPs. We demonstrate a method to determine the orientations of QWPs required for these transformations. The same state transformation technique is used to determine the Mueller matrix of an optical component and verified experimentally for free space.

\end{document}